# INDUCEMENT OF SPIN-PAIRING AND CORRELATED SEMI-METALLIC STATE IN MOTT- HUBBARD QUANTUM DOT ARRAY

**PARTHA GOSWAMI**[†]

*Deshbandhu College, University of Delhi, Kalkaji, New Delhi-110019,India*

**AVINASHI KAPOOR**

*Department of Electronic Science, University of Delhi South Campus, Delhi, India*

**Abstract.** We model a quantum dot-array (with one electron per dot) comprising of two (or more than two) coupled dots by an extended Hubbard Hamiltonian to investigate the role played by the inter-dot tunneling amplitude $t_d$ ,together with intra-dot (U) and inter-dot($U_1$) coulomb repulsions, in the singlet / triplet bound state formation and evolution of the system from the Mott-insulator-like state to a correlated semi-metallic state via charge-bond-order route. In the presence of magnetic field, $t_d$ is complex due to the appearance of Peierls phase factor. We introduce a short-ranged inter-dot capacitive coupling $U_0$, assumed to be non-zero for nearest-neighbor dots only, for the bound state analysis. The study indicates that, while for the tunable parameter $d = (2t_d/U_0)$ greater than unity only the possibility of the triplet bound state formation exists, for d less than one both triplet and singlet states are possible. The bound states are formed due to tunneling and capacitive dot-bondings with coulomb interactions $(U,U_1)$ playing marginal role. The interaction U, however, is found to play, together with complex $t_d$, an important role in the evolution of the double quantum dot system from the insulator-like state to that of a correlated semi-metallic state through charge-bond-ordering route.

## I. INTRODUCTION

It is possible to organize quantum dots in complex arrays, and these arrays possess features, which combine the properties of solid state and atomic physics [1,2]. In the case of strong Coulomb blockade, the number of electrons in a quantum dot (QD) is fixed, and one should discriminate between the dots with even and odd electron occupation. We consider here an array with dots belonging to the latter category only , viz. a system of N coupled dots with one electron per dot. This is modeled by a Hamiltonian (H) involving inter-dot tunneling parameter $t_d$, intra-dot and inter-dot coulomb repulsions $(U,U_1)$ and short-ranged inter-dot attraction $U_0$ to investigate the role played by $(t_d,U,U_0)$ in the singlet/triplet bound state formation. The parameter $t_d$ is complex when the magnetic field is assumed to be present; the interaction $U_0$ is essentially an electrostatic bonding. The approach to the problem involves setting up the Schrodinger equation in momentum space of the composite spin-paired object using the model Hamiltonian above without considering the single-particle terms. The integral equation obtained is solved in a manner similar to that in the well-known cooper pair problem. It is found that for the tunable parameter $\delta = (2t_d/U_0) < 1$ the possibility of singlet state formation exists. As regards triplet state, even an infinitely small value of $U_0$ is sufficient. The tunability of δ stems from the fact that $U_0$ could be changed by suitably tailoring the parameters, such as dot-size, inter-dot spacing, potential scenario ,etc. The coulomb interactions $(U,U_1)$ do not play significant role on this issue as long as the coulomb blockade condition (one electron per dot) is valid. For δ <1, both singlet and triplet states are possible. These point towards the fact that the system may act as the requisite structure for the basic gate operations of quantum computing. We have completely ignored spin-orbit coupling(SOC) in the dot system considered here; if SOC is introduced, the singlet state would mix with triplet state. Without spin-orbit coupling, the states of two electrons are either pure singlet or pure triplet. It may be mentioned that, several years ago, Loss and DiVincenzo [3,4] had mooted the proposal of the state-swap based on their investigation of a semi-phenomenological model Hamiltonian H=$J_{S-T}(t)\mathbf{S_1}\cdot\mathbf{S_2}$, where $\mathbf{S_1}$ and $\mathbf{S_2}$ are the spin-1/2 operators for the two localized electrons and the effective Heisenberg exchange splitting $J_{S-T}(t)$ is a function of time t. Moreover, in a quantum dot with two active levels for transport and an even number of electrons on the dot is known, experimentally as well as theoretically [5-7], to yield singlet-triplet crossover in the presence of magnetic field due to Kondo effect. The background discussed above provides the motivation behind the present communication where we wish to bring to the fore the fact that inter-dot tunneling and bonding in a Mott-Hubbard (MH) array could also be tuned to have possibilities of singlet-triplet and pure triplet bound states in the coulomb blockade regime. This also generates the hope of showing, in future, the possibility of Loss and DiVincenzo [3,4] type state-swap in a MH system with large state-swap-time ($T_{swap} \sim U/t_d$ and $U \gg t_d$ for a MH system) introducing the time-dependence explicitly in a more refined analysis .

---

[†] Corresponding author: e-mail: physicsgoswami@gmail.com;Tel:0091-129-243-9099.

We extend the model Hamiltonian considered above with the inclusion of the single-particle terms [8] involving explicit dependence on magnetic field in section III. The inter-dot tunneling amplitude is complex due to the appearance of the Peierls phase factor [8]. The purpose of the inclusion is to obtain single-dot(or single-particle) spectrum through which we wish to examine the evolution of a dot array from the insulator-like to the correlated semi-metal-like state relaxing slightly the MH condition $U \gg t_d$ in the coulomb blockade regime (CBR). The effect of magnetic field is directly involved here. In this regime the quantum dots representing a molecule is very weakly coupled to the source and drain which means that the charging energies U are still greater than the energies corresponding to contact, magnetic field and temperature induced broadening of the electron levels in the dot. We have, in fact, the modest aim to examine whether additional energy level can surface in the Mott-gap region of the MH-system aided by magnetic field and the charging energy U. We consider a minimal system - a double quantum dot(DQD) and corresponding Hamiltonian including Zeeman term etc., due to the presence of a magnetic field, for two coupled dots (double quantum dot(DQD)) to analyze the role played by the complex tunneling parameter together with coulomb interactions in the evolution of the system from the Mott-insulator-like state to a correlated semi-metallic state. We find that one possible route for the evolution corresponds to magnetic field aided charge-bond-ordering (CBO). We assume the tunneling event to be spin-conserving for simplicity and, therefore, introduce the Green's functions for each spin channel separately. The Green's functions are needed to compute the spectral function for the system under consideration where expectedly the magnetic field induced CBO peak manifestation would occur. The signature of this evolution can possibly be seen in many experiments, such as the one on the magnetic field dependence of the differential conductance.

The paper is organized as follows: In section II we investigate the interplay of capacitive and tunneling couplings in singlet and triplet bound state formation in a MH quantum dot-array (with one electron per dot) comprising of N ≥ 2 coupled dots. In section III we show that the interaction U, together with complex tunneling coupling $t_d$, plays an important role in the evolution of the double quantum dot system from the insulator-like state to that of a correlated metallic state through charge-bond-ordering route. The paper ends with a brief note on the possible uses of the investigation carried out.

## II. SINGLET AND TRIPLET STATES

A quantum dot array may be treated as "giant artificial molecule". It consists of N islands with confined fermions. The inter-dot coupling, comprising of ionic and covalent components, could be tuned by varying the width of tunnel channel between them. In fact, both capacitive and tunneling couplings exist between these islands. We extend these ideas to a D-dimensional(D ≥ 1) dot-array comprising of N coupled dots(with one fermion per dot) and visualize it as a conglomerate of tiny artificial atoms whose positions are specified by a site index. We next consider an extended Hubbard-like model for this dot-array in order to investigate the singlet/triplet bound state problem:

$$H = -\sum_{\langle ij \rangle \alpha \sigma} [t_d\, c^{\dagger}_{i\alpha\sigma} c_{j\alpha\sigma} + h.c] + U \sum_{i\alpha} n_{i\alpha\uparrow} n_{i\alpha\downarrow}$$

$$+ U_1 \sum_{\langle ij \rangle' \alpha} n_{i\alpha} n_{j\alpha} - \sum_{\langle ij \rangle \alpha \alpha'} U_0^{\alpha\alpha'}\, n_{i\alpha} n_{j\alpha'} \,. \qquad (1)$$

The Fermi operator $c^{\dagger}_{i\alpha,\sigma}$ creates a localized spin σ fermion at $i^{th}$ dot. The operators $n_{i\alpha\sigma} = c^{\dagger}_{i\alpha\sigma} c_{i\alpha\sigma}$ and $n_{i\alpha} = \sum_{\sigma} c^{\dagger}_{i\alpha\sigma} c_{i\alpha\sigma}$. Here i and j (which correspond to lattice sites in a Mott-Hubbard system) are indices locating dots in the cluster. The tunneling is assumed to be constrained to nearest neighbor (NN) dots ⟨ij⟩ ; $t_d$ is the complex tunneling amplitude of fermions, owing to presence of a magnetic field, between the NN dots and corresponds to the kinetic energy. The indices α,α' = {e,h}. We include the effect of short range interactions only: U is the intra-dot coulomb repulsion (This is assumed to be very large Coulomb repulsion energy which will have to be paid if a dot is occupied by two fermions) and $U_1$ is the inter-dot repulsion. Here ⟨ij⟩′corresponds to next-nearest-neighbor (NNN) dots. A tunneling coupling term involving NNN dots will be added later. $U_0^{\alpha\alpha'}$ is the capacitive coupling, assumed to be non-zero for NN dots only. The attraction essentially creates an electrostatic bond. Furthermore, the coupling may be assumed to be stronger for α ≠ α′ than that for α=α′ as a result of mutual attraction between electron and hole. The Hamiltonian (1) provides us a common plank to discuss the bound state formation for α = α′ (i.e. between electron-electron or hole-hole only) and α ≠ α' ( i.e. excitonic(e-h) bound state).

We presently consider two-fermion singlet bound state. With dot indices denoted by {l,m,…..}, the total wave function in the spin-singlet state can be written as

$$|\Psi\rangle = \sum_{l,m,\alpha,\alpha'} \Psi(l,m)(1/\sqrt{2}) \{\Phi_l^{\alpha} \Phi_m^{\alpha'} + \Phi_m^{\alpha} \Phi_l^{\alpha'}\} \chi^{\alpha\alpha'}_{s=0} |0\rangle \qquad (2)$$

$$\chi^{\alpha\alpha'}{}_{s=0} = (1/\sqrt{2})\,(\chi_{\alpha\uparrow}\,\chi_{\alpha'\downarrow} - \chi_{\alpha\downarrow}\,\chi_{\alpha'\uparrow}). \qquad (3)$$

Here $\Psi(l,m)$ is the tight-binding amplitude to describe the system; $\Phi$ and $\chi$, respectively, correspond to the spatial and spin parts of the wave function. In terms of the fermion creation and destruction operators $c^\dagger_{l\alpha\sigma}$ and $c_{m\alpha'\sigma}$ one can write $|\Psi> = \sum_{l,m} \Psi(l,m) |\Phi_s^{\alpha\alpha'}(l,m)>$ where

$$|\Phi_s^{\alpha\alpha'}(l,m)> = (1/\sqrt{2}) \sum_{\alpha\alpha'} \{ c^\dagger_{l\alpha\uparrow} c^\dagger_{m\alpha'\downarrow} - c^\dagger_{l\alpha\downarrow} c^\dagger_{m\alpha'\uparrow} \}|0>. \qquad (4)$$

We now write the Schrodinger equation for the composite spin-paired object as follows:

$$E |\Psi> = H \sum_{l,m} \Psi(l,m) |\Phi_s^{\alpha\alpha'}(l,m)> = \sum_{l,m} \Psi(l,m)\, H\, |\Phi_s^{\alpha\alpha'}(l,m)> \qquad (5)$$

$$E |\Psi> = \sum_{l,m} \Psi(l,m)\{ -\sum_{<ij>\alpha\sigma} [t_d\, c^\dagger_{i\alpha\sigma} c_{j\alpha\sigma} + h.c ] + U \sum_{i\alpha} n_{i\alpha\uparrow} n_{i\alpha\downarrow}$$
$$+ U_1 \sum_{<ij>'\alpha} n_{i\alpha} n_{j\alpha} - \sum_{<ij>\alpha\alpha'} U_0^{\alpha\alpha'}\, n_{i\alpha} n_{j\alpha'} \}|\Phi_s^{\alpha\alpha'}(l,m)> \qquad (6)$$

The equation for $\Psi(l,m)$ is obtained taking scalar product of both the sides of the equation above with $<\Phi_s^{\alpha\alpha'}(l',m')|$. Since $<\Phi_s^{\alpha\alpha'}(l',m')|\Psi> = \Psi(l,m)$ we have

$$E\,\Psi(l,m) = -\sum_j \{ t_d\,\Psi(j,m) + t_d^*\,\Psi(l,j) \} + U\,\Psi(l,m)\,\delta_{l,m}$$
$$- \sum_{\alpha\alpha',\check{z}} U_0^{\alpha\alpha'}\,\Psi(l,m)\,\delta_{l,m+\check{z}} + U_1 \sum_{\acute{z}} \Psi(l,m)\,\delta_{l,m+\acute{z}}. \qquad (7)$$

Since we are assuming that the dot-array is similar to a solid-state environment, we may define the Fourier transform(FT)

$$\Psi(k,q) = N^{-1} \sum_{l,m} \Psi(l,m)\,\exp\{i\,((k/2)+q).l + i\,((k/2)-q).m\}, \qquad (8)$$

where N is the number of dots in the array ( at half-filling the number L of quasi-particle like fermions will be equal to N), to go over to k- space (momentum space). Taking FT is being used here as more of a theoretical tool; it may or may not be physically meaningful. We use this for the simple reason that we find the solutions of integral equation corresponding to (7) easily tractable due to close similarity with the well-known Cooper pair problem [23]. We have included excitonic bound state consideration additionally this far, for electrons as well as exciton may act as the requisite structure for the basic gate operations of quantum computing.

As mentioned in section I the system under consideration being a dot array with one electron per dot, hereinafter we shall focus only on singlet and triplet electronic bound states with $U_0^{\alpha\alpha'}$ replaced by $U_0$ - a purely capacitive coupling. We introduce the sums $F_1(K) = N^{-1} \sum_k \Psi(K,k)$, $F_2(K,\check{z}) = N^{-1} \sum_k \Psi(K,k)\exp(-i\,\mathbf{k}.\check{\mathbf{z}})$ and $F_3(K,\acute{z}) = N^{-1} \sum_k \Psi(K,k)\exp(-i\,\mathbf{k}.\acute{\mathbf{z}})$ and set up three equations to determine them. Thereafter, we obtain a condition for the solutions of these equations to exist which we wish to cast in the well-known (Cooper pair problem) form $(-1/G(\varepsilon)) = N^{-1} \sum_k (\varepsilon - 2\varepsilon_k)^{-1}$. We need to look for the negative energy solution(s) of this equation to examine the two-electron bound state formation condition. To this end, we multiply Eq.(7) by $\{ \exp\{i ((k/2)+q).l+i ((k/2)-q).m\}$ and sum over all (l,m). After some algebra, we obtain from Eq.(7)

$$E \Psi(K,q) = \{ \varepsilon((K/2)+q) + \varepsilon^*((K/2)-q) \} \Psi(K,q) + (U/N)\sum_k \Psi(K,k)$$

$$- ( z_1 U_0 /2N ) \sum_k ( \Lambda_1 (q-k)+ \Lambda_1 (k-q) ) \Psi(K,k)$$

$$+ (z_2 U_1 /2N ) \sum_k ( \Lambda_2 (q-k)+ \Lambda_2 (k-q) ) \Psi(K,k) \quad (8)$$

where $\varepsilon(k) = -\sum_{\check{z}} t_d \exp(i\mathbf{k},\check{z})-\sum_{\acute{z}} t'_d \exp(i\mathbf{k},\acute{z})$, $\Lambda_1(k) = z_1^{-1}\sum_{\check{z}} \exp(i\mathbf{k}.\check{z})$, $\Lambda_2(k) = z_2^{-1} \sum_{\acute{z}} \exp(i\mathbf{k}.\acute{z})$, and $z_1$ is the number of nearest neighbor while $z_2$ is the number of next near neighbor (NNN). We have introduced a NNN hopping term with tunneling amplitude $t'_d$ above. Equation (8) can be rewritten in terms of the sums $F_1(K)$, $F_2(K)$ and $F_3(K)$ as

$$\Psi(K,q) = (\Gamma/D)$$

where
$\Gamma=[UF_1(K)-z_1U_0\Lambda_1^{av}(q)F_2(K)+z_2U_1\Lambda_2^{av}(q)F_3(K)]$, $D=[E-\varepsilon((K/2)+q)-\varepsilon^*((K/2)-q)]$, $\Lambda_1^{av}(q)=(1/2)(\Lambda_1(q)+\Lambda_1(-q))$, and $\Lambda_2^{av}(q)= (1/2) (\Lambda_2(q) + \Lambda_2(-q) )$. In view of these results, equations for $(F_1,F_2,F_3)$ can be written in the neat form

$$F_1( 1- U I_0) + (z_1U_0 I_1) F_2 + (- z_2U_1 J_1) F_3 = 0$$

$$F_1( - U I_1) + (1+z_1U_0 I_2) F_2 + (- z_2U_1 J_2) F_3 = 0$$

$$F_1( - U J_1) + (z_1U_0 J_2) F_2 + (1- z_2U_1 J_3) F_3 = 0 \quad (9)$$

where

$$I_n = N^{-1}\sum_q (\Lambda_1^{av}(q))^n / D, \quad J_1= N^{-1}\sum_q (\Lambda_2^{av}(q)) / D,$$

$$J_2 = N^{-1}\sum_q (\Lambda_1^{av}(q) \Lambda_2^{av}(q))/ D, \quad J_3 = N^{-1}\sum_q (\Lambda_2^{av}(q))^2 / D. \quad (10)$$

The system of equations (9) will have solutions for $F_j$ (j = 1,2,3) provided the determinant of the coefficients of $F_j$ in (9) is zero. We obtain

$$(1- U I_0) (1+z_1U_0I_2)(1- z_2U_1J_3)+ (1- U I_0) (z_1 z_2 U_0 U_1J_2^2) + (z_1 U_0 UI_1^2) (1- z_2U_1J_3)$$

$$+(2 I_1z_1 z_2 UU_0 U_1 J_1J_2) + (-z_2 U_1 UJ_1^2) (1+z_1U_0I_2) = 0. \quad (11)$$

This is the equation to investigate the two-dot(with one electron per dot) bound state formation. In conventional Cooper pair problem [23] the strength of the attractive interaction is maximum when K = 0, i.e. electron pairs with equal and opposite wave vectors. On a similar note, we assume here K =0.It is then easy to see from ref.8 that, in the weak magnetic field limit, one may write $D \approx [E + 2t_d \Lambda_1^{av}(q) + 2t'_d \Lambda_2^{av}(q)]$ and $I_0 = N^{-1} \sum_q (1/D)$. Now the integrals $I_0, I_1$ and $I_2$ can be rewritten in the form

$$I_0 = N^{-1} \sum_k (\varepsilon - 2\varepsilon_k)^{-1}, \quad \varepsilon = E + W_1 + W_2, \quad 2\varepsilon_k = W_1(1 - \Lambda_1^{av}(k)) + W_2(1 - \Lambda_2^{av}(k))$$

$$I_1 = W_1^{-1}(1 - E\, I_0 - W_2 J_1), \quad I_2 = -(E/W_1^2)(1 - E\, I_0) + (W_2/W_1)((E\, J_1/W_1) - J_2) \qquad (12)$$

where $W_1 = 2z_1 t_d$ and $W_2 = 2z_2 t'_d$. As regards the evaluation of other integrals $J_1, J_2$, etc., we carry it out treating $\Lambda_2$, which corresponds to next nearest neighbors, to be q-independent. We find that the integrals, viz. $J_1, J_2, J_3$ and $I_n$ (n=1,2,..), are expressible in terms of $I_0$.

It is tedious but straightforward to show that, under the approximations stated above, (11) may be written in the conventional form as

$$N^{-1} \sum_k (\varepsilon - 2\varepsilon_k)^{-1} = [\{(b+c)f_1(E) + bcB\}/\{(b+c)f_2(E) + bc\, f_1(E) + 2bc(E+c)B - b\, cU\, B\}] \qquad (13)$$

where $f_1(E) = [1 - a - B(E-U)]$, $f_2(E) = [U(1+a+BE) - BE^2 - 2aE - ac]$, $a = [(z_1 U_0 W_2 \Lambda_2)/W_1^2]$, $b = (z_2 U_1 \Lambda_2^2)$, $c = W_2 \Lambda_2$, and $B = [(z_1 U_0)/W_1^2]$. Equation (13) has been written down retaining all possible terms corresponding to NN as well as NNN. The latter ones are found to complicate the analysis without any trade-off of capturing some important piece of physics. Therefore, in what follows we approximate (13) as

$$f(\varepsilon) = (-1/G(\varepsilon)) \approx N^{-1} \sum_k (\varepsilon - 2\varepsilon_k)^{-1} \qquad (14)$$

where

$$-G^{-1}(\varepsilon) = -\{W_1^2/(\eta z_1 UU_0)\} [\{1 + (z_1 UU_0(1-\eta)/2W_1^2)\}/(\varepsilon - \varepsilon_{1P})]$$

$$+ \{W_1^2/(\eta z_1 UU_0)\}[\{1 + (z_1 UU_0(1+\eta)/2W_1^2)\}/(\varepsilon - \varepsilon_{2P})], \qquad (15)$$

$$\varepsilon_{1P} = \{(1/2) U(1+\eta) + W_1\}, \quad \varepsilon_{2P} = \{(1/2) U(1-\eta) + W_1\}, \quad \eta = \{1 + (4 W_1^2/z_1 UU_0)\}^{1/2}. \qquad (16)$$

The function $f(\varepsilon)$ has simple poles at $(\varepsilon_{1P}, \varepsilon_{2P})$ and it intersects the energy axis at $\varepsilon = W_1 + U + (W_1^2/z_1 U_0)$. The function has local minimum at $\varepsilon_1 = W_1 + U$ and local maximum at $\varepsilon_2 = W_1 + (U/2)(1 + \eta^2)$. We find $f(\varepsilon_1) = U^{-1}$, $f(\varepsilon_2) = (\eta^2 U)^{-1}$, and $\{-f(\varepsilon_{1P} \pm 0^+), f(\varepsilon_{2P} \pm 0^+)\} \to \pm \infty$. Moreover, we find $f(0) = U^{-2}[\{z_1 UU_0 W_1^{-1}(1 + U W_1^{-1}) + U\}/\{1 - z_1 U_0 (W_1^{-1} + U^{-1})\}]$ which implies that $f(0) > 0$ provided $z_1 U_0 (W_1^{-1} + U^{-1}) < 1$ and it is less than zero when the opposite is true. Similarly, if $2 z_1 U_0 (W_1^{-1} + 2U^{-1}) < 1$, $f(-W_1) > 0$ otherwise it is less than zero. In view of these conclusions it is not difficult to see (Figs.1 and 2) that, for $\varepsilon_{2P} < 0$ which implies $U_0 < 2t_d$, Eq.(14) has no solution whereas, for $\varepsilon_{2P} > 0$ which implies $U_0 > 2t_d$ (i.e. quite strong capacitive coupling), a negative energy solution is possible. In Figs.1 and 2 we have plotted dimensionless energy $(\varepsilon/U_0)$ along x-axis and both $f(\varepsilon/U_0)$ and $(N)^{-1} \sum_k U_0(\varepsilon - 2\varepsilon_k)^{-1}$ along y-axis. The bound state energy values are given by the abscissa of the points which are intersection of the curve $(N)^{-1} \sum_k U_0(\varepsilon - 2\varepsilon_k)^{-1}$ (uneven-toothed curve) with the function $f(\varepsilon/U_0)$ (with simple poles at $(\varepsilon_{1P}/U_0, \varepsilon_{2P}/U_0)$) on the negative side of $(\varepsilon/U_0)$. The conclusion from above is that in the dot-array system comprising dots with odd electron occupation the formation of singlet bound state is possible provided the inter-dot capacitive bonding is stronger than the inter-dot tunneling. We notice that the coulomb repulsions $(U,U_1)$ do not play significant role as long as size of a quantum dot is compatible with the assumption that there is one electron per dot. We now summarize below in brief a similar analysis carried out for triplet bound state.

For the triplet bound state, the wave functions, in terms of the tight-binding amplitude $\Psi(l,m)$, have the following form:

$$|\Psi_1\rangle = \sum_{l,m} \Psi(l,m)\, c^{\dagger}_{l\uparrow} c^{\dagger}_{m\uparrow} |0\rangle, \quad |\Psi_0\rangle = \sum_{l,m} \Psi(l,m)\{ c^{\dagger}_{l\uparrow} c^{\dagger}_{m\downarrow} + c^{\dagger}_{l\downarrow} c^{\dagger}_{m\uparrow} \}|0\rangle,$$

$$|\Psi_{-1}\rangle = \sum_{l,m} \Psi(l,m)\, c^{\dagger}_{l\downarrow} c^{\dagger}_{m\downarrow} |0\rangle. \tag{17}$$

The equation for $\Psi(l,m)$ is same as Eq.(7) except the fact that the anti-symmetry of $\Psi(l,m)$ demands that $\Psi(m,m)$ should be zero now. We thus have

$$E\,\Psi(l,m) = -\sum_j \{t_d\,\Psi(j,m) + t_d^{*}\Psi(l,j)\} - \sum_{\check{z}} U_0 \Psi(l,m)\,\delta_{l,m+\check{z}} + U_1 \sum_{\acute{z}} \Psi(l,m)\,\delta_{l,m+\acute{z}}. \tag{18}$$

In the same way, the counterpart of Eq.(8) may be written as

$$E\,\Psi(K,q) = \{\varepsilon((K/2)+q) + \varepsilon^{*}((K/2)-q)\}\Psi(K,q) - (z_1 U_0/2N)\sum_k (\Lambda_1(q-k)$$

$$+ \Lambda_1(k-q))\Psi(K,k) + (z_2 U_1/2N)\sum_k (\Lambda_2(q-k) + \Lambda_2(k-q))\Psi(K,k). \tag{19}$$

After some amount of algebra, in the weak magnetic field approximation, ignoring the NNN term as before we find for triplet bound state a cooper-pair like equation

$$(-1/2U_0) \approx F(\varepsilon);\quad F(\varepsilon) = N^{-1} \sum_q \sin^2 qa/(\varepsilon - 2\varepsilon_q), \tag{20}$$

Here 'a' is separation between NN dots. Equation (20) corresponds to triply degenerate solution(s) which have been sought graphically in Fig.3. In this figure we have plotted dimensionless energy $(\varepsilon/U_0)$ along x-axis and $F(\varepsilon/U_0)$ along y-axis. The bound state energy values are given by the abscissas of the points which are intersection of the horizontal line $(-1/2)$ with the function $F(\varepsilon/U_0)$ on the negative side of $(\varepsilon/U_0)$. The lowest among such values of $(\varepsilon/U_0)$ corresponds to ground state of triplet cooper-like-pair. We encounter here no restriction involving the capacitive coupling $(U_0)$ and the tunneling coupling$(t_d)$. It is thus clear from above that for $U_0 < 2t_d$ only a triplet bound state is possible, whereas for $U_0 > 2t_d$ triplet as well as singlet states are possible. The finding generates the hope of showing, in future, the possibility of Loss and DiVincenzo [3,4] type state-swap in a MH system with large state-swap-time introducing the time-dependence explicitly in a more sophisticated theoretical formulation.

## III. DOUBLE QUANTUM DOT

The coherent tunneling coupling and the capacitive (electrostatic) coupling are the well-known coupling mechanisms [6-9] for all possible dot-arrays including the simplest of such systems, viz. a double quantum dot(DQD). The electrostatic coupling constants $U_{ij}$, in the limit $C \ll C_g$ where C is the capacitive coupling between the $i^{th}$ dot and neighboring dots and $C_g$ the capacitance of a dot with relative to external gates, are given by $U_{ij} \sim U(C/C_g)^{|i-j|}$ [10, 11]. The tunneling coupling $t_d$, on the other hand, delocalizes the electron wave-function spreading over the neighboring dots and splitting energy states into singlet and triplet in the absence of SOP. As it is clear from above, the electrostatic bonding energy is smaller than the covalent bonding energy only when $C \ll C_g$ and U is not too large compared to $t_d$. In this limit, we investigate here the possibility of inducement of semi-metal like state in a DQD by considering complex tunneling coupling. We have the modest aim to examine whether additional level can surface in the Mott-gap region of DQD system aided by magnetic field and the charging energy U via charge-bond-order route. It may be relevant to mention here that somewhat similar systems, such as artificial heterostructures and superlattices based on two different MH insulators, are known [12,13] to exhibit the possibility of tuning the carrier density by the electric field effect and a superconducting transition too at very low temperature.

The simplest of all possible dot-arrays, as stated above, is the double quantum dot (DQD) first realized experimentally in planar geometry. We consider a minimal Hamiltonian for DQD in the presence of a magnetic field assumingly weak, to analyze the role played by the tunneling amplitude $t_d$ ($t_d = t \exp(i\varphi)$ where $\varphi$ is the Peierls phase factor [8]) together with the coulomb repulsion U, in the evolution of the system from the Mott insulator like state to a correlated semi-metallic state:

$$H = -\mu' \sum_\sigma (c^\dagger_{1\sigma} c_{1\sigma} + c^\dagger_{2\sigma} c_{2\sigma}) - \sum_\sigma [t_d\ c^\dagger_{1\sigma} c_{2\sigma} + h.c ]$$

$$+ U(n_{1\uparrow} n_{1\downarrow} + n_{2\uparrow} n_{2\downarrow}),\ \mu' \equiv \mu - \Delta, \qquad (17)$$

$$\Delta = \hbar [(\omega_c/2)^2 + \omega_0^2]^{1/2} + (-1)^\sigma (g \mu_B B /2). \qquad (18)$$

The first term in (17) is the intra-dot term which includes the chemical potential $\mu$, the cyclotron frequency $\omega_c$, the Zeeman term ($g \mu_B B /2$) and the confinement effect (e.g. for harmonic oscillator confinement potential in the absence of a magnetic field $\Delta = \hbar \omega_0$ with $\omega_0$ being essentially a harmonic oscillator frequency). We consider a single spin-split level per dot, and set $\sigma = 1,2$ in (18) to correspond to the spin up/down lowest confined quantum dot level. We neglect corrections arising from single-particle level crossing [8] assuming the magnetic field to be weak. It may be noted that since we have also assumed here the electrostatic bonding energy to be smaller than the covalent bonding energy (which holds true when $C \ll C_g$ and U is not too large compared to $t_d$), we need not include the term ($-\sum_{\sigma\sigma'} U_0 n_{1\sigma} n_{2\sigma'}$) in (17). We are interested here in single-particle spectrum, which is given by poles of the Fourier coefficient of the temperature function

$g_{\alpha\alpha'\sigma}(\tau) = -\langle T\{c_{\alpha\sigma}(\tau) c^{\dagger}_{\alpha'\sigma}(0)\}\rangle$ (where $\alpha, \alpha' = 1,2$ and T is the time-ordering operator which arranges other operators from right to left in the ascending order of time $\tau$), to fulfill the objective stated in section I. In what follows we summarize the technique adopted and the important steps in the calculation.

The presence of atomic and molecular like spectra of quantum dot array calls for appropriate theoretical tools of investigation. Since the exact diagonalization method [14,15] has its own limitations and Hartree-Fock method [16,17] suffers from sizeable systematic errors, one may treat the energy states of the array using Density Functional Theory[18] with the exchange interaction involvement [19,20] explicitly. However, the spin-dependent properties are not exactly the focal point of the investigation at present. Besides, the strong coulomb repulsion U is to be handled properly. We, therefore, prefer the equation of motion (EOM) method which is known to be reliable in the coulomb blockade regime and also qualitatively correct for the Kondo regime [21,22].

Since the Hamiltonian H is not diagonal one can write down the equations for the operators $\{c_{1\sigma}(\tau), c_{1\sigma}(\tau) n_{1\sigma}(\tau),\ldots\}$, where the time evolution an operator O is given by

$$O(\tau) = \exp(H\tau) O \exp(-H\tau), \qquad (18)$$

in Hubbard approximation to ensure that the thermal averages, such as $g_{\alpha\alpha'\sigma}(\tau)$ etc., are determined in a consistent manner. Upon retaining terms of the type $\langle n_{\alpha\sigma}\rangle (= \langle c^{\dagger}_{\alpha\sigma} c_{\alpha\sigma}\rangle)$, and $\langle n_{\alpha\alpha'\sigma}\rangle (= \langle c^{\dagger}_{\alpha\sigma} c_{\alpha'\sigma}\rangle)$ in EOMs we find

$$(\partial/\partial\tau) g_{11\uparrow}(\tau) = \mu' g_{11\uparrow}(\tau) - U \Gamma_{11\uparrow}(\tau) - t_d g_{21\uparrow}(\tau) - \delta(\tau), \qquad (19)$$

$$(\partial/\partial\tau) \Gamma_{11\uparrow}(\tau) \approx -(U-\mu') \Gamma_{11\uparrow}(\tau) - t_d g_{21\uparrow}(\tau) \langle n_{1\downarrow}\rangle$$
$$- t_d g_{11\uparrow}(\tau) (\langle c^{\dagger}_{1\downarrow} c_{2\downarrow}\rangle - \langle c^{\dagger}_{2\downarrow} c_{1\downarrow}\rangle) - \langle n_{1\downarrow}\rangle \delta(\tau), \qquad (20)$$

$$(\partial/\partial\tau) g_{21\uparrow}(\tau) = \mu' g_{21\uparrow}(\tau) - U \Gamma_{21\uparrow}(\tau) - t_d g_{11\uparrow}(\tau), \qquad (21)$$

$$(\partial/\partial\tau) \Gamma_{21\uparrow}(\tau) \approx -(U-\mu') \Gamma_{21\uparrow}(\tau) - t_d g_{11\uparrow}(\tau) \langle n_{2\downarrow}\rangle$$
$$+ t_d g_{21\uparrow}(\tau) (\langle c^{\dagger}_{1\downarrow} c_{2\downarrow}\rangle - \langle c^{\dagger}_{2\downarrow} c_{1\downarrow}\rangle), \qquad (22)$$

where $\Gamma_{11\uparrow}(\tau) = -\langle T\{c_{1\uparrow}(\tau) n_{1\downarrow}(\tau) c^{\dagger}_{1\uparrow}(0)\}\rangle$, $\Gamma_{21\uparrow}(\tau) = -\langle T\{c_{2\uparrow}(\tau) n_{2\downarrow}(\tau) c^{\dagger}_{1\uparrow}(0)\}\rangle$, and $g_{21\sigma}(\tau) = -\langle T\{c_{2\sigma}(\tau) c^{\dagger}_{1\sigma}(0)\}\rangle$. The quantity $\mu' \equiv \mu - \hbar[(\omega_c/2)^2 + \omega_0^2]^{1/2} + (g\mu_B B/2)$. In writing these equations we have ignored correlation between opposite spins. The Fourier coefficients of the thermal averages above are the Matsubara propagators $\{g_{11\uparrow}(z), \Gamma_{11\uparrow}(z), g_{21\uparrow}(z), \Gamma_{21\uparrow}(z)\}$ where $z = [(2n+1)\pi i/\beta]$ with $n = 0, \pm 1, \pm 2, \ldots$. With the aid of Eqs.(19)-(22) one obtains the following equations of these propagators:

$$(z + \mu') g_{11\uparrow}(z) + (-U) \Gamma_{11\uparrow}(z) + (-t_d) g_{21\uparrow}(z) = 1,$$

$$-t_d (\langle c^\dagger_{1\downarrow} c_{2\downarrow} \rangle - \langle c^\dagger_{2\downarrow} c_{1\downarrow} \rangle) g_{11\uparrow}(z) + (z - (U - \mu')) \Gamma_{11\uparrow}(z) + (-t_d) \langle n_{1\downarrow}\rangle g_{21\uparrow}(z) = \langle n_{1\downarrow}\rangle,$$

$$(-t_d) g_{11\uparrow}(z) + (z + \mu') g_{21\uparrow}(z) + (-U) \Gamma_{21\uparrow}(z) = 0,$$

$$(-t_d) \langle n_{2\downarrow}\rangle g_{11\uparrow}(z) + t_d (\langle c^\dagger_{1\downarrow} c_{2\downarrow}\rangle - \langle c^\dagger_{2\downarrow} c_{1\downarrow}\rangle) g_{21\uparrow}(z) + (z - (U - \mu')) \Gamma_{21\uparrow}(z) = 0. \quad (23)$$

It must be noted that the EOMs for the other relevant Matsubara propagators { $g_{22\uparrow}(z)$, $\Gamma_{22\uparrow}(z)$, $g_{12\uparrow}(z)$, $\Gamma_{12\uparrow}(z)$ } can be obtained easily from above provided the interchange $1\leftrightarrow 2$ and the replacement $t_d = t \exp(i\varphi) \rightarrow t \exp(-i\varphi)$ are made. For the propagators of spin-down channel, however, $\mu'$ has to be replaced by $\mu'' \equiv \mu - \hbar [(\omega_c/2)^2 + \omega_0^2]^{1/2} - (g \mu_B B /2)$. In writing the equations above we have completely ignored the correlation between intra-dot and inter-dot opposite spins thereby keeping the possibility of spin-density-wave-like and spin-bond-order-like states formation out of the purview of the present communication.

Suppose, to begin with, we completely ignore the inter-dot correlations in Eqs.(23). Then it is easy to see that the Matsubara propagator $g_{11\uparrow}(z)$ is given by

$$g^{-1}_{11\uparrow}(z) = g^{(0)-1}_{11\uparrow}(z) - \Sigma^{(0)}_{12}(z) \quad (24)$$

where

$$g^{(0)}_{11\uparrow}(z) = [\{(1 - \langle n_{1\downarrow}\rangle)/(z + \mu')\} + \{\langle n_{1\downarrow}\rangle/(z + \mu' - U)\}], \quad (25)$$

$$\Sigma^{(0)}_{12}(z) = t_d^2 [\{(1 - \langle n_{2\downarrow}\rangle)/(z + \mu')\} + \{\langle n_{2\downarrow}\rangle/(z + \mu' - U)\}]. \quad (26)$$

This grossly approximate expression of the propagator $g_{11\uparrow}(z)$ clearly indicates that the tunneling coupling $t_d$ (or the magnetic field) has the potential to alter the single-particle spectrum in a fundamental way. Taking cue from here we have solved Eqs.(23) for all the propagators involved without making any approximation. In particular, for $g_{11\uparrow}(z)$, we obtain in the closed form

$$g^{-1}_{11\uparrow}(z) = g^{(1)-1}_{11\uparrow}(z) - \Sigma^{(\text{eff})}_{12}(z). \quad (27)$$

where the self-energy part $\Sigma^{(\text{eff})}_{12}(z) = g^{(1)-1}_{11\uparrow}(z) g^{(0)}_{11\uparrow}(z) \Sigma^{(0)}_{12}(z)$ and the propagator

$$g^{(1)}_{11\uparrow}(z) = [A_1(z + \mu')^{-1} - B_1 U(z + \mu')^{-2} + A_2(z + \mu' - U)^{-1} - B_2 U(z + \mu' - U)^{-2}$$

$$+ A_3 (z + \mu' - U(1 - \langle n_{2\downarrow}\rangle))^{-1}]. \quad (28)$$

The coefficients $\{A_1, A_2, \ldots\}$ are given by

$$A_1 = [(1 - \langle n_{1\downarrow}\rangle) - (2t_d/U)(1 - \langle n_{1\downarrow}\rangle)(1 - \langle n_{2\downarrow}\rangle)^{-1}(\langle n_{12\downarrow}\rangle - \langle n_{21\downarrow}\rangle)], \quad (29)$$

$$A_2 = [\langle n_{1\downarrow}\rangle + (2t_d/U) \langle n_{1\downarrow}\rangle \langle n_{2\downarrow}\rangle^{-1}(\langle n_{12\downarrow}\rangle - \langle n_{21\downarrow}\rangle)], \quad (30)$$

$$A_3 = (2t_d/U)(\langle n_{1\downarrow}\rangle - \langle n_{2\downarrow}\rangle)(1 - \langle n_{2\downarrow}\rangle)^{-1}(\langle n_{12\downarrow}\rangle - \langle n_{21\downarrow}\rangle) \langle n_{2\downarrow}\rangle^{-1}, \quad (31)$$

$$B_1 = (t_d / U)(1 - \langle n_{1\downarrow}\rangle)(\langle n_{12\downarrow}\rangle - \langle n_{21\downarrow}\rangle), \quad B_2 = (t_d / U)\langle n_{1\downarrow}\rangle(\langle n_{12\downarrow}\rangle - \langle n_{21\downarrow}\rangle). \quad (32)$$

Since we have already assumed U to be not too large compared to $t_d$, $(\langle n_{1\downarrow}\rangle, \langle n_{2\downarrow}\rangle) \neq 1$. We shall now consider the terms $(\langle n_{12\downarrow}\rangle - \langle n_{21\downarrow}\rangle)$ and $(\langle n_{1\downarrow}\rangle - \langle n_{2\downarrow}\rangle)$ in (31) and (32). The aim is to show that these terms are non-zero only in the presence of a magnetic field. As it is clear from (28) to (32), this leads to emergence of an intermediate energy state $\varepsilon = (U(1 - \langle n_{2\downarrow}\rangle) - \mu')$ in the Mott-gap-like region between the states $\varepsilon = -\mu'$ and $\varepsilon = (U - \mu')$.

The Fourier coefficient $g_{21\uparrow}(z)$, from (23), is given by $g_{21\uparrow}(z) = (\Delta_3 / \Delta)$ where $\Delta$ is the determinant of the coefficients of $\{g_{11\uparrow}(z), \Gamma_{11\uparrow}(z), g_{21\uparrow}(z), \Gamma_{21\uparrow}(z)\}$ in (23) and

$$\Delta_3 = t_d (z + \mu' - U)^2 + U t_d (\langle n_{1\downarrow}\rangle + \langle n_{2\downarrow}\rangle)(z + \mu' - U) + U^2 t_d \langle n_{1\downarrow}\rangle \langle n_{2\downarrow}\rangle. \quad (33)$$

For $g_{12\uparrow}(z)$, $t_d (= t \exp(i\varphi))$ above will get replaced by $t_d^* (= t \exp(-i\varphi))$. This implies that the term $(\langle n_{12\downarrow}\rangle - \langle n_{21\downarrow}\rangle)$, which corresponds to $\sum_n \exp(-z\, 0^+)(g_{21\uparrow}(z) - g_{12\uparrow}(z))$, will involve $(t \sin \varphi)$. Thus the charge-bond-order like states in a DQD system, in the presence of a magnetic field, induce an intermediate energy state in the Mott-gap region. An interesting question is what would be the effect on this intermediate state when the capacitive coupling involving term is also included in the analysis. We wish to address this issue in future. Presently, our task is to examine the spectral function to find out whether the state is sufficiently long-lived one.

The single-dot spectral function (SF) in the spin-up channel is given by $A_{11\uparrow}(\omega) = (-\pi^{-1}) \mathrm{Im}\, G^R_{11\uparrow}(\omega)$, where $G^R(\omega)$ is a retarded Green's function given by

$$G^R_{11\uparrow}(\omega) = {}_{-\infty}\!\!\int^{\infty} (d\omega'/2\pi)\{\zeta_{11\uparrow}(\omega')/(\omega - \omega' + i\, 0^+)\} \quad (34)$$

and $\zeta_{11\uparrow}(\omega) = -i\{g_{11\uparrow}(z)|_{z=\omega-i0+} - g_{11\uparrow}(z)|_{z=\omega+i0+}\}$. Similarly, spectral function in the spin-down channel can be obtained replacing $\mu'$ by $\mu''$. In view of (28)-(34) we find that spin-up channel SF is given by

$$A_{11\uparrow}(\omega) \approx \{A_1 - 2t_d (1 - \langle n_{1\downarrow}\rangle)(\langle n_{12\downarrow}\rangle - \langle n_{21\downarrow}\rangle) P((\omega + \mu')^{-1})\} \delta(\omega + \mu')$$

$$+ \{A_2 - 2t_d \langle n_{1\downarrow}\rangle(\langle n_{12\downarrow}\rangle - \langle n_{21\downarrow}\rangle) P((\omega + \mu' - U)^{-1})\} \delta(\omega + \mu' - U)$$

$$+ A_3\, \delta(\omega + \mu' - U(1 - \langle n_{2\downarrow}\rangle)). \quad (35)$$

Here P represents a Cauchy's principal value. Since $A_{11\uparrow}(\omega)$ is a bunch of delta functions (a Fermi-liquid-like feature) all the states are long-lived ones. An increase in magnetic flux through the system increases $(t \sin \varphi)$ alluded to above which, in turn, is expected to give rise to taller peak for the intermediate energy state $\varepsilon = (U(1 - \langle n_{2\downarrow}\rangle) - \mu')$ due to increased $A_3$ (see also Eq.(31)); it brings about decrease in the peak-height corresponding to the energy states $\varepsilon = -\mu'$ and $\varepsilon = (U - \mu')$. It must be mentioned here that more than a decade ago Ugajin [24-26] had shown the possibility of external field driven metal-insulator transition in quantum dot super-lattices . Our investigation seems to be a corroboration of this reporting for a Mott-Hubbard DQD.

Our findings above are depicted through a sketch in Figure 4.The double-peaked curve corresponds to a spectral function obtainable from Eqs.(24) and (25) while the curve with three peaks corresponds to Eq.(35). The figure depicts the emergence of correlated semi-metal-like feature from insulator-like state due to the presence of a magnetic field. This change indicates the usefulness of correlated DQD as a sensor. The effect has the potential of modulating excitations locally and can be applied to new field-effect devices. Also, the bound state consideration for electrons discussed in section II may be useful in the search of a requisite structure for the basic gate operations of quantum computing.

# FIGURE CAPTIONS:

**FIGURE 1.** In this figure we have plotted dimensionless energy ($\varepsilon/U_0$) along x-axis and both $f(\varepsilon/U_0)$ and $(N)^{-1} \sum_k U_0(\varepsilon-2\varepsilon_k)^{-1}$ along y-axis. The singlet bound state energy values are given by the abscissa of the points which are intersection of the curve $(N)^{-1} \sum_k U_0(\varepsilon-2\varepsilon_k)^{-1}$ (uneven-toothed curve) with the function $f(\varepsilon/U_0)$ (with simple poles at ($\varepsilon_{1P}/U_0$, $\varepsilon_{2P}/U_0$)) on the negative side of ($\varepsilon/U_0$). Here $\varepsilon_{2P} > 0$ which implies $U_0 > 2t_d$ (i.e. the capacitive coupling is quite strong in comparison to the tunneling coupling).

**FIGURE 2.** In this figure we have plotted dimensionless energy ($\varepsilon/U_0$) along x-axis and both $f(\varepsilon/U_0)$ and $(N)^{-1} \sum_k U_0(\varepsilon-2\varepsilon_k)^{-1}$ along y-axis. Here $\varepsilon_{2P} < 0$ which implies $U_0 < 2t_d$ (i.e. the capacitive coupling is quite weak in comparison with the tunneling coupling). The plot shows non-feasibility of singlet bound state formation in this case.

**FIGURE 3.** In this figure we have plotted dimensionless energy ($\varepsilon/U_0$) along x-axis and $F(\varepsilon/U_0)$ along y-axis. The bound state energy values are given by the abscissas of the points which are intersection of the horizontal line $(-1/2)$ with the function $F(\varepsilon/U_0)$ on the negative side of ($\varepsilon/U_0$). The lowest among such values of ($\varepsilon/U_0$) corresponds to ground state of triplet cooper-like-pair. We encounter here no restriction involving the capacitive coupling ($U_0$) and the tunneling coupling ($t_d$).

**FIGURE 4.** This figure depicts a qualitative plot of $A_{11\uparrow}(\omega)$, given by Eq.(35) approximately, as a function of $\omega$. The double-peaked curve corresponds to a spectral function obtainable from Eqs.(24) and (25) while the curve with three peaks corresponds to Eq.(35). The figure represents the emergence of correlated semi-metal-like feature from insulator-like-state for a DQD due to the presence of a magnetic field.

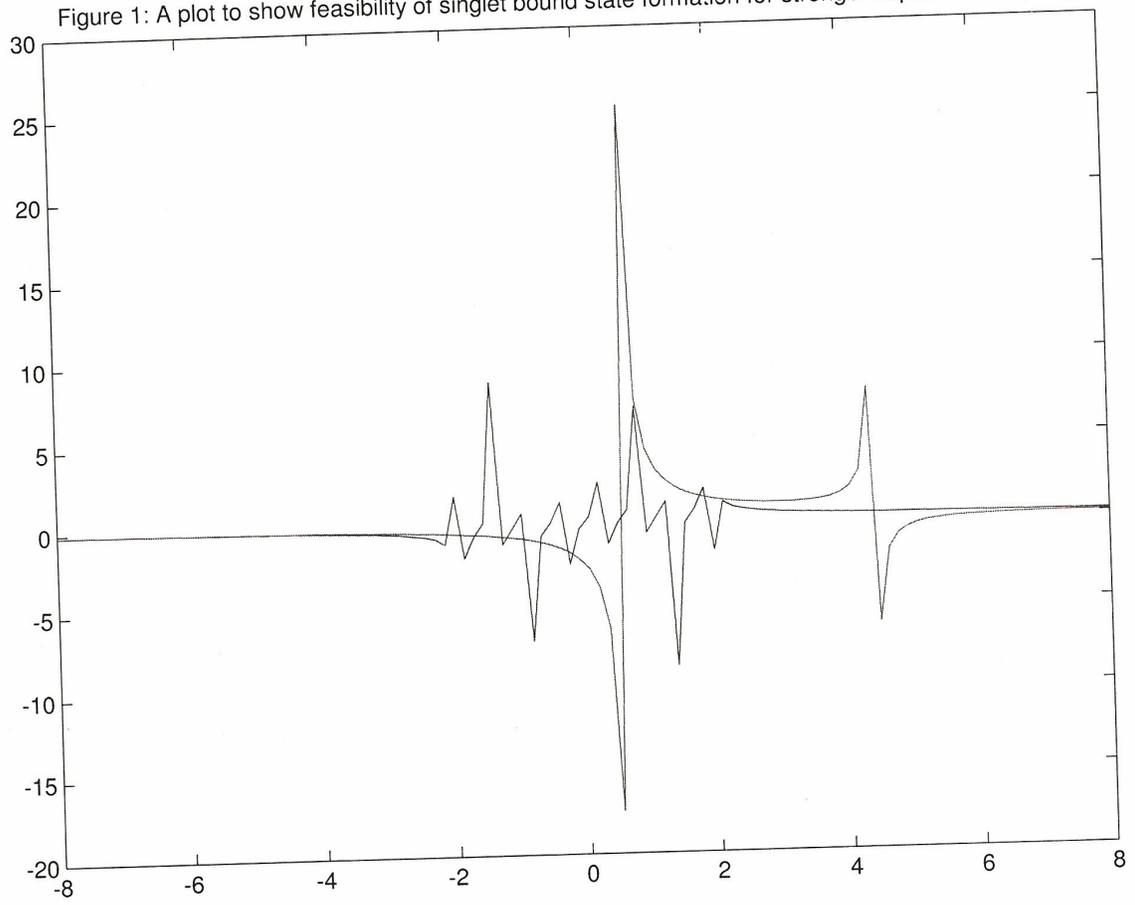

Figure 1: A plot to show feasibility of singlet bound state formation for stronger capacitive coupling.

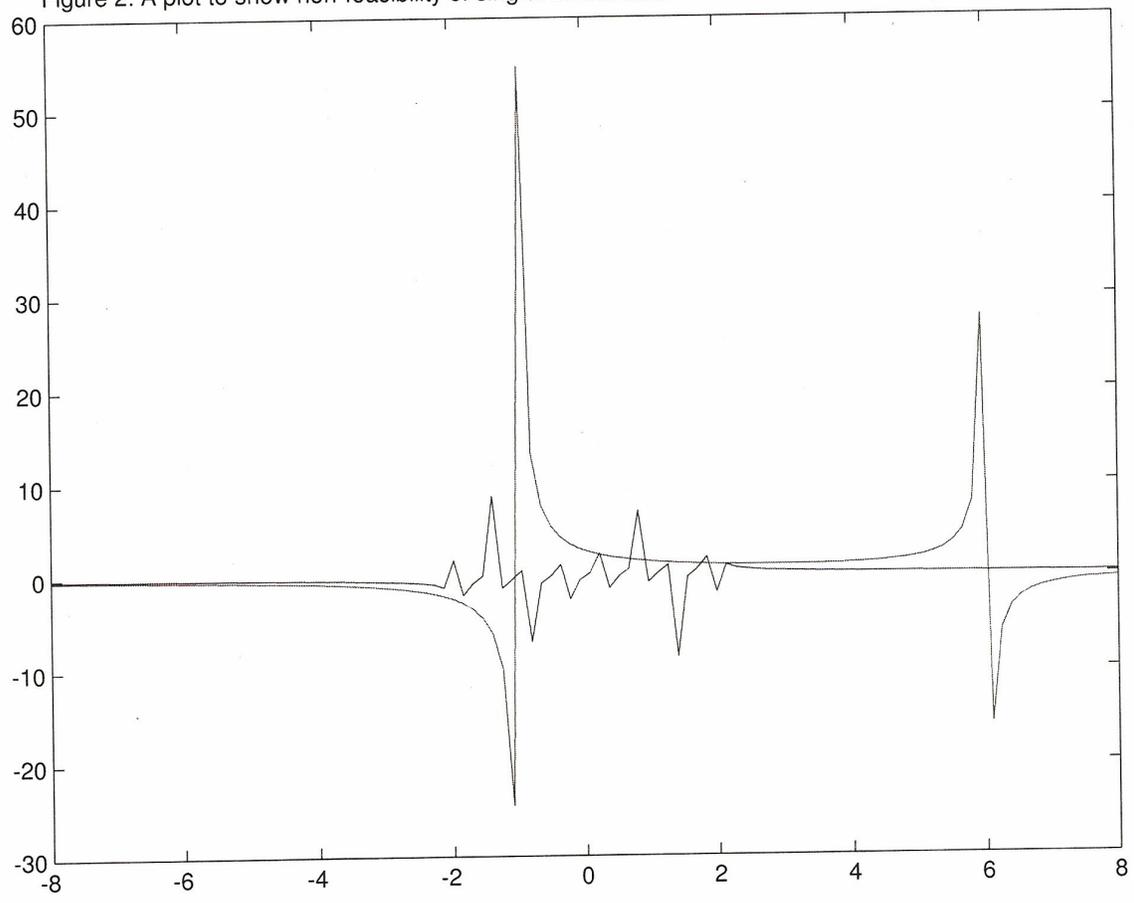
Figure 2: A plot to show non-feasibility of singlet bound state formation for weaker capacitive coupling.

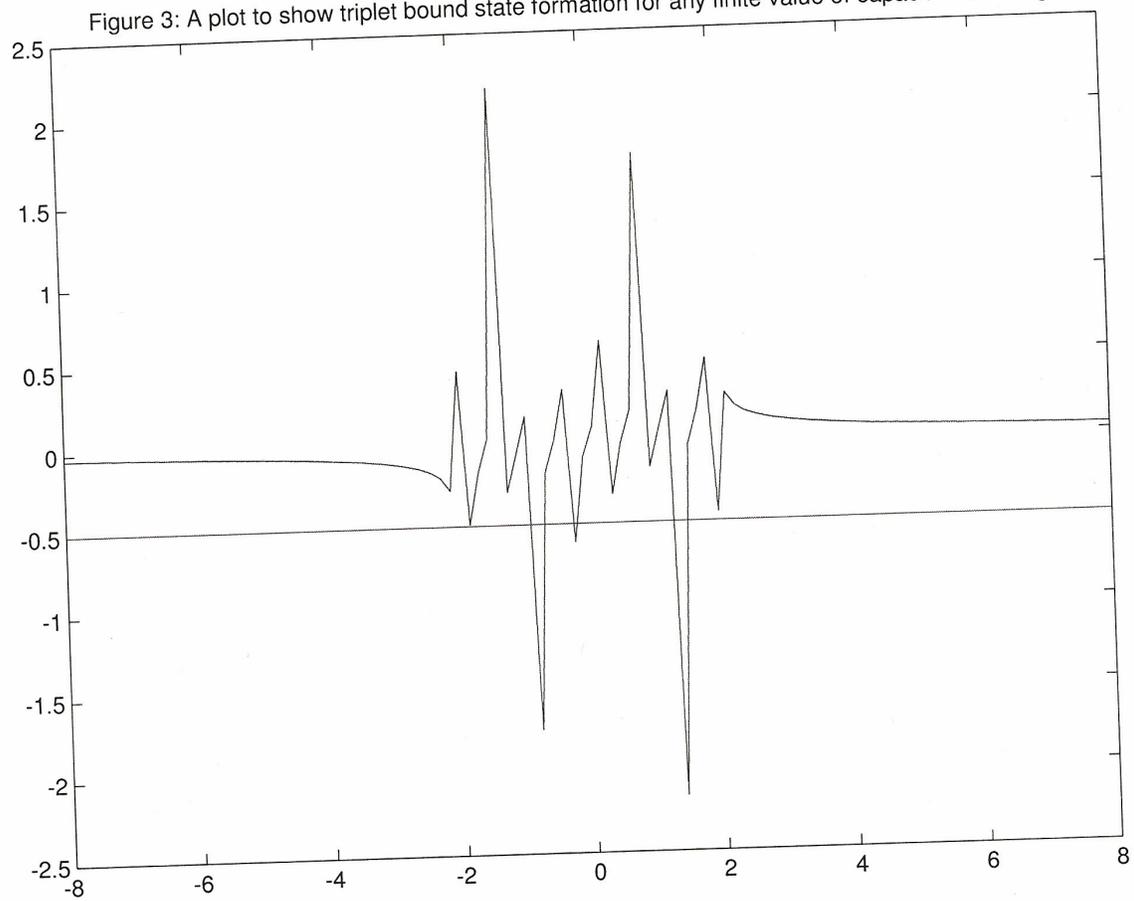

Figure 3: A plot to show triplet bound state formation for any finite value of capacitive bonding.

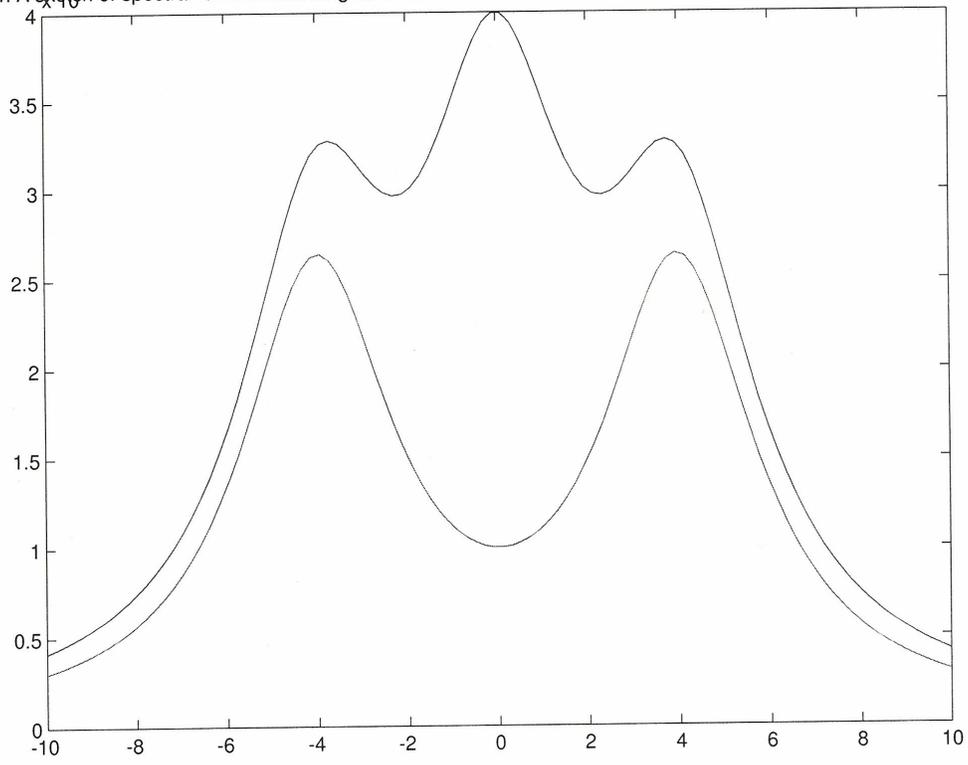

Figure 4: A sketch of spectral function showing its evolution to a correlated metal like state due to application of magnetic field.